\documentclass[12pt]{article}
\def\={\!=\!}

\usepackage{amsmath,amssymb,amscd}

\usepackage[usenames]{color}

\def\s{\sigma}
\def\oDelta{\Delta}
\def\la{\lambda}
\def\w{\widetilde}
\def\p{\partial}

\def\={\!=\!}

\def\dom{{\rm dom }}

\def\|{{\Vert}}

\def\a{\alpha}

\newcommand{\R}{\mathbb R}

\newcommand{\Z}{\mathbb Z}

\def\T{{\cal T}}

\def\d{{\partial}}
\def\dd{{\delta}}
\def\M{{\cal S}}

\newcommand{\noi}{\noindent}

\def\skip{\vspace{2mm}}
\newcount\u
\def\Remark{\skip\noi{{\bf{Remark \number\u.}}} \advance\u by 1}

 \title{Myopic equilibria, the spanning property and subgame bundles}
\author {R. Simon, S. Spie\.z, H. Toru\'nczyk}
\begin{document}
\maketitle
\thispagestyle{empty}
\vfill

\noi London School of Economics\newline
Department of Mathematics\newline
Houghton Street\newline
 London WC2A 2AE\newline

 \noi Institute of Mathematics\newline
 Polish Academy of Sciences\newline
 {\'S}niadeckich 8, 00--656 Warszawa

\vfill

\date{}

\setcounter{page}{-1}
  \noi

\newpage
\vskip2cm
\thispagestyle{empty}
\noi

\vskip1cm

{\bf Abstract:}
\ \ {For a set-valued function $F$ on a 
compact subset $W$ of a manifold, spanning is a topological property that implies 
that $F(x)\ne \emptyset$ for interior points $x$ of $W$.} A myopic equilibrium  applies when for each action there is a payoff  whose functional
value is not necessarily affine in the strategy space.
We show that if the payoffs
satisfy the spanning property, then
there exist a myopic equilibrium (though not necessarily a
Nash equilibrium).
Furthermore, given a parametrized collection of games and the spanning
property to
the structure of payoffs in that collection, the resulting
myopic equilibria and their payoffs
have the spanning property with respect to that parametrization.
This is a far reaching extension of the Kohberg-Mertens Structure Theorem.
There are at least four useful applications,
when payoffs are
exogenous to a finite game tree (for example a finitely repeated game followed by
an infinitely repeated game), when one wants to understand
a game strategically entirely with behaviour strategies,
when one wants to extends the subgame concept to subsets
of a game tree that are known in common, and for evolutionary game theory.
 {The proofs involve new topological results asserting that spanning is preserved by relevant operations on set-valued functions.}

\vskip1cm

\noi

\vskip2cm

\newpage

\section{Introduction}

 Conventionally with games payoffs are multilinear functions of the mixed strategies, meaning that with finitely many pure strategies the payoffs  can be represented by multidimensional matrices. One can move away from a multilinear relationship between mixed strategies and payoffs and still get a Nash equilibrium. What counts is that the best reply correspondence of each player in mixed strategies is a convex set, allowing one to apply fixed point theory. If however the relationship between the mixed strategies of a player and her payoffs is merely continuous, there may not be a Nash equilibrium. For example, if the payoff function for a player is convex in her mixed strategies, the best reply correspondence can generate disjoint sets, and the result may be no Nash equilibria.

This lack of a Nash equilibrium can be rectified by expanding the concept of what is a payoff for a player. Instead of associating a payoff to a mixed strategy, one can associate a payoff to each pure strategy or action
 that the player can use. This broadening of what is a payoff allows one to define a new kind of equilibrium, called a myopic equilibrium. 
A myopic equilibrium is defined, see (Simon, Spie\.z and Toru\'nczyk (2020)), to be a strategy profile such that  each action used with positive probability by a player  gives the maximum payoff possible from the use of any action of that player.
 In the context of multilinear payoffs,  a myopic equilibrium is the same as a Nash equilibrium.
 { But even when the payoff functions are concave, the myopic equilibria may be far from the Nash equilibria, as we will see in Example 1 below.}

In the present paper, we consider  parametrized relationships between the mixed strategies and the payoffs, especially when that relationship satisfies
the spanning property (Simon, Spie\.z, Toru\'nczyk 2002). We show that if the payoffs relative to a parametrization have the spanning property, then the resulting parametrized myopic equilibria also have the spanning property.

This theory has multiple new applications.

First, myopic equilibria are equivalent to equilibria used in evolutionary game theory. The move away from a continuous relationship between mixed strategies and payoffs should broaden the scope of evolutionary game theory.

Second, with extensive form games, the relationship between mixed strategies and behaviour strategies can be seen in a new light. Behaviour strategies are difficult to work with because they lack the multilinear relationship between the strategies and payoffs. Mixed strategies are difficult to work with because the dimension of the strategy space is much higher than with behaviour strategies. The translation between behaviour strategies and mixed strategies, accomplished by Kuhn (1953), allowed one to go between a strategic structure that was easier to comprehend (behaviour strateges) and one for which a Nash equilibrium exists (mixed strategies). But with myopic equilibria and the spanning property, one can prove the existence of Nash equilibrium entirely through the behaviour strategies.

Third, we can  understand what is a subgame in a new way. Subgames are linked closely to the concept of common knowledge, that the players know in
common that they have landed in the subgame. A problem arises if the subset of starting points is not a single point; a single subgame is not defined,
 rather a collection of subgame parametrized by the probability distributions on the starting points of this subset.
 But with the spanning property and myopic equilibria applied to this parametrization, we can understand a plurality of starting points as a kind of subgame.

Fourth, we can prove the existence of equilibria for some new forms of  games through induction, using the conservation of the spanning property in the relationship between the strategies, payoffs, and equilibria.

The following example demonstrates some of the uses of our theory.

\vskip.2cm

\noi {\bf Example 1:} \vskip.2cm

\noi
Consider the following game of three players.
Player One chooses between two states, $X$ and $Y$ and
Players Two and Three are not informed
of the outcome of this choice. Then Players
Two and Three play a simultaneous move zero-sum game between each other (so that
this game is equivalent to a simultaneous move three person game).
Player Two has the actions $L$ and $R$ and Player Three has the
actions $l$ and $r$.
The payoff matrices for Players Two and Three are
the following:
$$\mbox { State X } \quad \quad \begin{matrix}& l & r \cr L& 9 & 1 \cr R&
1 & 1\end{matrix}
\quad \quad \mbox { State Y } \quad \quad
\begin{matrix}& l & r \cr L & 1 & 1
\cr R& 1 & 9\end{matrix}$$ where the entries in the matrices
represent the payoffs
that Player Three must pay to Player Two.
Player One gets a payoff of $0$ for all combinations except for
$R$ and $l$ in State $X$ and $L$ and $r$ in State $Y$.
For the $Rl$ combination in State $X$ she gets $1+s $ for some
very small $s> 0$ and in State Y she gets $1$ for the
$Lr$ combination. What are the Nash equilibria of this game and how does it relate to myopic equilibria and a parametrized collection of subgames?

\vskip.2cm

\noi We start with the strategies for Player One. Let $p\in [0,1]$ be
the probability for choosing the state $X$. Assuming that
Players Two and Three know this probability $p$, (though not assumed
mathematically
in the definition of an equilibrium, this
can be assumed in its application),
their choices are informed by the payoff matrix created by
the mixture of $p$ times the matrix for State $X$ plus $1-p$
times the matrix for State $Y$:
$$\begin{matrix} & & \beta & 1-\beta \cr \cr \alpha & & 8p+1 & 1
\cr 1-\alpha& & 1 & 9-8p \end{matrix}$$ \newline
-- for the sake of analysis we added $\alpha$ and $\beta$,
$\beta$ for
the probability that Player Three will choose $l$
and $\alpha$ for the probability that Player Two will choose $L$.
If $p$ lies in the interior $(0,1)$ there are
no optimal pure strategies for either player. Assuming indifference between
the two actions, we get the formulas
\newline $(1+8p) \beta + 1-\beta = \beta + (1-\beta) (9-8p)$ and \newline
$(1+8p) \alpha + 1 -\alpha = \alpha+ (1-\alpha) (9-8p) $, \newline which are
symmetric in $\beta$ and $\alpha$ and solve to $\beta=\alpha=1-p$.
Now consider Player One's payoff. Her choice for $X$ results in a payoff
of $(1+s) \beta (1-\alpha) = (1+s) (1-p) p$.
Her choice for $Y$ results in a payoff of
$\alpha (1-\beta ) = (1-p) p$. As long as $p$ is not
equal to either $0$ or $1$,
the choice for $X$ is superior. It follows that there are two
types of equilibria in the three player game. In one type,
Player One chooses $X$ with certainty, Player Three
chooses $r$ with certainty, and Player Two chooses any value for
$\alpha$. In the other type, Player One chooses $Y$ with certainty,
Player Three chooses $l$ with certainty, and Player Two chooses any value
for $\alpha$. In both cases Player One receives a payoff of $0$. Note
however that if $s=0$ then all values for $p$ can be combined with
the appropriate above strategies for players Two and Three to define
Nash equilibria of the three player game.

\vskip.2cm

\noi Now lets think of this situation
as a one player game in which Player One
chooses some $p\in [0,1]$ (representing the probability for $X$) and
what follows is no longer part
of that game but nevertheless the payoff to Player One must
correspond to what the other players
would do in equilibrium. The choice for $X$, given $p$, gives a payoff
of $(1+s) (1-p)p$. The choice for $Y$, given $p$, gives a payoff
of $(1-p)p$.
This means that the choice for $p$ yields an expected payoff
of $f(p) : = (1+s) (1-p) p^2 + (1-p)^2 p$.
To optimize the payoff we must take the derivative
$f'(p) = 1+ (2s-2) p -3sp^2$ and set it to zero, which solves to
$p= \frac { s-1 + \sqrt {1+s +s^2}} {3s}$. For $s$ very small
the solution calls for $p$ very close to $\frac 12$, with an expected
payoff very close to $\frac 14$. Finding the
Nash equilibrium of a one player game
is a problem of optimization and as long as the payoff function
is continuous there will be at least one optimal strategy, and if
the payoff function is strictly concave there will be a unique optimal strategy.
Indeed the payoff function of this game is strictly
concave, as the second derivative of $f$ is
$2s-2 -6sp$, which is always negative as long as $s$ is less than $2$.
However the concept of Nash equilibrium is not the right one
for understanding
the three person game broken into two parts,
the first part being that of Player One's choice and the
second part collection of
 games between Players Two and Three. It corresponds
to the situation where Player One can commit
herself to a mixed strategy and maximizes the payoff accordingly.
But what if Player One ``takes back control'' to
choose the action giving the highest payoff?
By doing so she will destroy the equilibrium property of the one player game and after it is re-established through the myopic  equilibrium concept
she will have a worse payoff, though  one relevant to the three player game. This has similarity to the myth of
Odysseus and the Sirens, where Odysseus optimizes by binding himself.

\vskip.2cm

\noi Conventionally the game played between Players Two and Three is not
considered a subgame because it does not involve
common knowledge of a single state, even
though the players do know in common that their
choices pertain to a subset of
size two. This ``subgame'' is really a continuum of subgames, one for each
probability distribution on this subset of size 2.
\vskip.2cm

\noi We generalize the concept of subgame to
subsets of vertices in a game tree known in common, and
call it a {\em subgame bundle}. These subsets are
closed with respect to actions and the
knowledge of the players. Some requirement of
common knowledge is needed on the subsets that can define a subgame
bundle,
because otherwise a player could be forced
to play the same in two locations belonging to different subgames.

\vskip.2cm

\noi A problem with generalizing the concept of a subgame to
subsets
is that the set of equilibria, as a function of the
probability distributions on the subset of starting positions,
in general cannot be approximated by continuous functions.
The above Example 1 used a zero-sum game played between the remaining players,
whose equilibria as a function of distributions will be an upper-semi-continous and non-empty
convex valued corrspondence. However in the more general context
of non-zero-sum games the equlibrium payoff correspondence
will not have this nice property. \vskip.2cm

\noi
We know from
the Structure Theorem of Kohlberg and Mertens (1986)
that there will be a topological structure to the equilibrium
correspondence, a
homotopic relation that implies the  spanning property.
 There are two advantages to working with the spanning property over
the homotopic property of Kohlberg and Mertens.
First the spanning property is more general.
With repeated games of incomplete information
on one side the equilibrium payoff correspondence
as a function of the probability distributions
has the spanning property (Simon, Spie\.z, Toru\'nczyk 2002), but
there is no reason to believe that it has the homotopic property.
Second, there is no indication yet that the homotopic property is
robust with
respect to composition, though we prove that the spanning property
has such a robustness.

\vskip.2cm

\noi The myopic equilibrium concept is not entirely new. The mathematical
structure behind a myopic equilibrium is identical to that of
a ``Nash equilibria of population densities'' from evolutionary
game theory. Though the mathematical formalities are the same,
the concepts are very different. With evolutionary game theory, there is a
continuum of animals belonging to a species.
Each species has a variety of types and
a distribution of
types is equivalent mathematically to a mixed strategy.
What we call a
myopic equilibrium is in the context of evolutionary game
theory a kind of Nash
equilibrium because each individual animal seeks to maximize
a utility independent of
the species as a whole.
The distribution of types will influence the payoffs for each
type, and that introduces a potentially non-affine
structure to those payoffs.

\vskip.2cm

\noi The rest of this paper is organized as follows. In the second section we {provide the necessary topological apparatus and prove new results about the spanning property, giving conditions under which it is
preserved by operations on correspondences like taking intersections, taking cartesian products or taking weighted   sums.
In the third section} we define ``game bushes'', ``game bundles'',  ``subgame bushes'', and
``subgame bundles'' {and use results from \S 2 to show}
the robustness of the spanning property with respect to myopic equilibria, that
if the payoffs have the spanning property
then the equilibrium payoffs of the game bundle
has the spanning property. In the fourth
section we consider subgames, factor games, and a new concept of
subgame perfection.
Finally we consider related questions and problems.


\vskip1cm

\section{The spanning property of correspondences} %

Let $X$ and $Y$ be metrizable spaces. By a {\sl correspondence} $F: X\to Y$ we mean here any compact subset of
$X\times Y$. Hence the same correspondence may also be denoted as $F\subset X\times Y$ and we often switch from one notation to the other; however, the use of $F: X\to Y$ displays the asymmetric  role which $X$ and $Y$ play below.
Given such an $F$ and a subset $W$ of $X$ we call $F\cap (W\times Y)$ the {\sl{restriction}} of $F: X\to Y$ to $W$ and denote it $F|W$. The image of $F|W$ under the projection to $Y$ is denoted by $F(W)$.
We also write $F(x):=F(\{x\}) \mbox{ for } x\in X$ and  $\dom (F):= \{ x \in X: F(x)\ne \emptyset\}$, the {\sl{domain}} of $F$. The usual challenge we encounter is to show that a given point $x\in X$ is in $\dom (F)$, i.e, that $F(x)\ne \emptyset$.

It is convenient to note that if $F: X\to Y$ and $G: Y\to Z$ are correspondences, then the formula $H(x)=G(F(x))$ defines a correspondence, i.e., $\bigcup _{x\in X}\{x\}\times H(x)$ is a compact subset of $X\times Z$. We write $G\circ F$ for $H$.

\medskip 

\noi Until Theorem 1 below we let $M$ be a compact, connected $d$-manifold with boundary $\dd M$ (possibly empty) and $W$ be a compact subset of $M$. By $\dd W$ we denote the boundary of the interior of~$W$ with respect to $M$.  Important to us is that there exists a well-defined element of the \v Cech homology group $H_{d}(W, \dd W;\Z_2)$, denoted here by $[W]$, with the following properties:

\medskip
\noi {\bf Fact 1.} {\it a) {If $W$ is a compact, connected $d$--manifold} then $[W]$ is the $\Z _2$--fundamental class of $W$, the unique non-zero element of the \v Cech homology group $H_{d}(W, \dd W;\Z_2)$.

b) With $\partial [W]\in H(\dd W; \Z _2)$ standing  for the image of $[W]$ under the boundary operator, one has that the  image of $\partial [W]$ in $H (W\setminus \{w\}; \Z _2)$ under the inclusion--induced map is non-zero, for each point $w$ in the interior~of~$W$.}

\medskip A proof with  $M$ a sphere is given in (Simon, Spie\.z, Toru\'nczyk 2002); using the orientablity  of $M$ over $\Z _2$ a generalization causes no difficulty.
Of course, $[W]$ depends on the ambient manifold $M$, e.g. it is null when $W$ is a disk embedded in the sphere $S^3$, but non-null if this disk is being considered as a subset of $S^2$.

\medskip
We say that a correspondence $F : W\to Y$ has {\sl{property $\M$}} if {$[W]$ lies in the image of the projection--induced homomorphism $H(F, F|\dd W; \Z _2) \to H(W, \dd W;\Z _2)$ of the relative \v{C}ech homology groups}
with $\Z _2$--coefficients. If $F$ is given as, say, $F\subset W\times Y_1\times Y_2$ and we wish to consider it as a correspondence from $W$ to $Y_1\times Y_2$ (rather than e.g. from $Y_1$ to $W\times Y_2$), then above we say that $F$ has {\sl{property $\M$ for $W$}}, and similarly when there are more or two factors. Here, $\M$ is an abbreviation for ''spanning'', addressing to Fact 2a) below:

\medskip
\noi {\bf Fact 2.} {\it a) If $F: W\to Y$ has property $\M$ then the domain of $F$ contains the closure of the interior of $W$ relative to the ambient manifold $M$.

b) If $W'\subset W$ are compact subsets of $M$ and $F: W\to Y$ has property $\M$ then $F|W'$ hast it either.

c) Let $W$ and $W'$ be compact subsets of $M$ and let correspondences $F: W\to Y$ and $F': W'\to Y$ coincide on $W\cap W'$. If $F' $ takes values in singletons only, then  $F\cup F'$ has property $\M$ for $W\cup W'$.

d) Suppose a correspondence $F: W\to Y$ is such that for each neighbourhood
$U$ of $\dd W$ in $M$ and each neighborhood $V$
of $F$ in $W\times Y$ there exists a compact set  $G\subset V$ with  property $\M$ for a compact set
$D$ satisfying $\dd D\subset U$ and $D\subset W \cup U$. Then, $F$ has property $\M$ for $W$.
}

\medskip Claim a) above follows from Fact 1b), and for the other ones see (Simon, Spie\.z, Toru\'nczyk 2002) (again, with $M$ a sphere).

\bigskip
\noi {\bf Theorem 1.} {\it Let $W, X, Y$ be compact 
manifolds and $F, G\subset W\times X\times Y$ be correspondences  with property $\M$ for $W\times X$ and
for $W\times Y$, respectively, such that the projections $F\to Y$ and $G\to X$ are null--homotopic.
Then, $F\cap G$ has property $\M$ for $W$.}

\medskip\noi {\it Proof.} 1). Let's first assume none of $W, X, Y$ has a boundary. Let $\a\in H (F)$  be a homology class mapped to $[W\times X]$ under the homomorphism induced by the projection to $W\times X$.  The projection of $F$ to $Y$ being homotopic to a constant (say, $y_o$) one has $\a= [W\times X\times \{y_o\}]$ in $H(W\times X\times Y)$ -- by which we mean that the images of these two classes under the inclusion-induced homomorphisms coincide.

Similarly, there exists a class $\beta \in H(G)$ mapped to $[W\times Y]$ under the homomorphism induced by the projection of $G$ to $W\times Y$ and such that for some $x_o\in X$ one has $\beta=[W\times \{x_o\}\times Y]$ in $H(W\times X\times Y)$.

Below, we'll be using the properties of the ''intersection pairing'' described in (Dold, 1972, \S VIII.13). So, there is a well-defined element $\a\bullet \beta$ of $H(F\cap G)$.
In $H(W\times X\times Y)$ this element is equal to  $[W\times X\times \{y_o\}]\bullet [ W\times \{x_o\}\times Y]$, by what said earlier. The  last product is  however  equal to $[ W\times \{(x_o, y_o)\}]$, for the manifolds $X\times \{y_o\}$ and $\{x_o\}\times Y$ intersect transversally at $(x_o, y_o)$. And since $ [ W\times \{(x_o, y_o)\}]$ is mapped to $[W]$ under the homomorphism induced by the projection to $W$, so is $\a\bullet \beta$. Thus $F\cap G$ has property $\M$~for~$W$.

2). Let now $\dd W=\emptyset $ and $\dd Y\ne \emptyset $. We then consider $Y$ as a subset of $\w Y:=Y\cup _{\dd Y} Y'$,  {a union of two copies $Y$ and $Y'$ of $Y$ intersecting along $\dd Y$. Let $h$ be a homeomorphism of $Y$ onto $Y'$ which restricts to identity on $\p Y$, and let $G':=({\rm id} _{W \times X}\times h)(G)$ and $\w G:=G\cup G'$.

We first prove that $\w G$ has property $\M$ for $W\times \w Y$. To this end we take an  $\a\in H (G, G|(W\times \dd Y))$ which witnesses property $\M$ of $G$ (i.e., is mapped to $[W\times Y]$ by the projection--induced homomorphism), set $\a ':=h_* (\a)$
and let $\w \a\in H(\w G, \w G|(\d W\times Y);\Z _2)$ be the image of $\a \oplus \a'$
in the relative Mayer-Vietoris sequence for $(\w G, \w G|(\d W\times Y)), (\w G', \w G'|(\d W\times Y'))$.
 We readily see that $\p \w \a=0$, so $\w \a $ can be considered as an element of $H( \w G)$. Also, the projection-induced homomorphism maps $\w \a$ to the image of $[W\times Y]\oplus [W\times Y']$ under an analogous Mayer--Vietoris  sequence, 
 and this image is equal to $[W\times \w Y]$. This shows that $\w G$ indeed has property $\M$ for $W\times \w Y$. 
(We skipped a technical point dealing with why are all the pairs involved excisive.)

Moreover, the projection of $\w G$ to $X$ factors through the projection $G\to X$ and hence is null-homotopic. The analogous property of the projection $ F\to \w Y$ is even more obvious.
Since $\dd \w Y=\emptyset$,  if additionally $\dd X=\emptyset $ then $F\cap \w G$ has property $\M$ for~$W$, by~1) above. However, $F\cap \w G=F\cap G$ and thus the assertion is true if $\dd W=\dd X=\emptyset$ (or if $\dd W=\dd Y=\emptyset$, by symmetry).

3). The assertion also holds true if both $X$ and $Y$ do have a boundary but~$W$ doesn't. This is so because then the reasoning of~2) still applies once the reference to~1) gets replaced by one to the conclusion of 2). (No additional assumption on~$X$ is being made this time.)

4). Finally, if $\dd W\ne \emptyset $ then we replace $W$ by $\w W:=W\cup _{\dd W}W$, and~$F$ and {$G$ by $\w F:=F\circ (q\times {\rm id}_{X\times Y})$ and $\w G:=G\circ (q\times {\rm id}_{X\times Y})$ respectively, where $q: \w W\to W$ is a retraction. As in 2) above we infer that 
$\w F$ has property $\M$ for $\w W \times X$ and $\w G$ has it for $\w W\times Y$. Hence,  the correspondence $\w F \cap \w G$ has property $\M$ for $\w W$, by ~1) to~3) above. Since on~$W$ it restricts to $F\cap G$, the desired conclusion follows from Fact 2b).~\hfill $\Box$}
\medskip 

\noindent{\bf Addendum to Theorem 1.}{\it The conclusion of Theorem 1 holds also true if~$X$ and $Y$ are as before and~$W$ is any compact subset of a compact, connected ambient manifold $M$ that is PL or is of dimension neither 4 nor 5.}

\medskip 

\noindent{Proof.}   The interior of $W$ is the union of an increasing sequence $(W_n)$ of compact manifolds with boundaries; see (Kirby and Siebenmann (1977), p.~108). By Theorem 1 and {Fact 2b)}, for each $n$ the appropriate restriction of $F\cap G$ has property $\M$ for $W_n$,  whence {Fact 2d)} applies. \hfill $\Box$

\bigskip
\noi {\bf Theorem 2.} {\it Let $W, X, Y $ be as in Theorem 1 or in the Addendum to it.
Furthemore, let $\Phi: W\to X$ and $\Psi: X\to Y$ be correspondences with property $\M$ such that the projections $\Phi \to X$ and $\Psi \to Y$ are null--homotopic.  Then, $\Psi\circ \Phi: W\to Y$ has property $\M$.
}

\medskip

\noindent {\it Proof.}
Let $F:=\Phi\times Y$ and $G:= W\times \Psi $.
Then $F, G\subset W\times X\times Y$ are correspondences  with property $\M$ for $W\times Y$ and
for $W\times X$, respectively, and such that the projections $F\to X$ and $G\to Y$ are null--homotopic.

We have $F\cap G=\{(w,x,y) ~|~ w\in W, x\in \Phi (w), y\in \Psi (x)\}$ and the projection of $F\cap G$ to $W$ is equal to the  projection of $F\cap G$ along~$X$ to $\Psi \cap \Phi$, followed by the projection of $\Psi\cap \Phi$ to $W$. Since $F\cap G$ has property $\M$ for $W$ (by Theorem~1 and the Addendum), so does $\Psi \cap \Phi$.
\hfill $\Box$

\vskip.2cm
\noindent{\bf Remark.}  {The assumption in Theorems 1 and 2 that the corresponding maps be null-homotopic (i.e., homotopic to constant ones) is automatically satisfied if $X$ and $Y$ are simplices, as in the applications in  \S 3 and \S4 below.}

\bigskip
\noindent
{\bf Theorem 3.}
{\it
Let $W$ be a compact top--dimensional subset of a PL-manifold and $F_i: W\to Y_i~ (i=1, \ldots  , l)$  be correspondences with property $\M$. Then, this property is also enjoyed by the correspondence $F: W \to \prod _i Y_i$ , defined by the formula $F(x)=\prod _iF_i(x)$.
\medskip

\noindent
{\it Proof.} }
We may assume all the $Y_i$'s to be normed linear spaces (for they can be embedded into such). For each  $ i$ let $V_i$ be an open neighborhood of $F_i$ in  $X \times Y_i$. We
 assume this neighbourhood is triangulated consistently with the PL-structure of $W\times prod _i Y_i$. By the spanning property of $F_i$, for $d=\dim (W)$ there exists a $d$--chain $z_i$  in $V_i$, with boundary in $\dd W \times Y_i$ and such that its class is mapped to $[W]$ by the projection--induced homomorphism.

By general position we may assume that with respect to some triangulation $\T$ of $W$  each $d$-simplex of $z_i$ projects injectively onto some simplex~of~$\T$.

For each tuple  $(\sigma_1, \ldots, \sigma_l)$  of $d$-simplexes $\sigma_i \in z_i$  which project onto the same simplex  of $\T$ we  define a ''diagonal'' $k$-simplex in $X\times \prod _iY_i$ by :
$$
\Delta(\sigma_1, \ldots, \sigma_l):= \{ (x,y_1,\dots, y_l)~ |~ (x,y_i) \in \sigma_i  \mbox{ for each} \  i \, \} \, .
$$
Then, we let $z$ denote the chain in $X \times \prod _i V_i$ which is the sum of $\Delta(\sigma_1, \ldots, \sigma_l)$'s  over all tuples  $(\sigma_i)_{i=1}^l$  as above.
One can observe that $z$ is  a $d$-chain  in $(W, \p W) \times \prod _i V_i$  and its class is mapped to $[W]$ by the projection--induced homomorphism .
Thus,  $F$ has the spanning property by {Fact 2d)}.    \hfill $\Box$

\bigskip
\noindent
{\bf  Corollary 1.}
{\it In Theorem 3 assume that $Y_i=\R ^A$ for each $i$, where $A$ is a finite set.
Then, $\sum _i F_i: W\to \R ^A$ defined by the formula $(\sum _iF_i)(x)=\{y_1+\dots +y_l~|~y_i\in F_i(x)\mbox{ for all }i\}$ is a correspondence with property $\M$.
}

\bigskip
\noindent
{\it Proof.} Let $Y\subset (\R ^A)^l$ be a closed ball containing $F(W)$, where $F(x)=\prod _i F_i(x)\subset (\R ^A)^l$ for $x\in W$. Obviously, $\sum _i F_i=f\circ F$, where $f(y)=\sum _iy_i$ for $y=(y_i)_{i=1}^l\in Y$. Hence, it remains to apply Theorems 2 and 3.
\hfill $\Box$

\bigskip \noindent
Given $F\subset X\times \R ^A$ and a function $\la : X\to \R$ we define $\la F\subset X\times \R ^A$ by the formula:
$$(\la F)(x)=\{\la (x)y: y\in F(x)\}\mbox{ for }x\in X.$$

\medskip
\noindent
{\bf  Corollary 2.} {\it Let $W$ be a compact PL--manifold and for $ i=1,2, \ldots , l$  let it be given a continuous function $\lambda_i: W \to \R$, a function $f_i : W \to Y_i$ which is continuous off $\la _i^{-1}(0)$, and a correspondence $G_i: Y_i \to \R^A$  with property $\M$. Then the correspondence
$\sum _i\lambda _i(G_i\circ f_i)$ has property $\M$.
}

\bigskip
\noindent
{\it Proof.} By Corollary 1 it suffices to consider the case where $l=1$; we hence write $\la , f$ and $G$ for $\la _1, f_1$ and $G_1$, respectively. We may assume that $\la \ne 0$, say $\sup |\la | (W)>1$. With $K_n:=\{x\in X~|~|\la (x)|
> 1/ n\}$ it follows from Theorem 2 and Fact 2b) that $G\circ f|K_n$ has property $\M$, whence $H_n:=\la G\circ f | K_n$ has it either. Let $K$ and $H$ denote the closures of  $\bigcup _n K_n$ and of  $\bigcup _n H_n$, respectively. Using boundedness of $G(W)$ we note that $H$ is compact and $H=(\la G\circ f)|K$; also, from {Fact 2d)} we infer easily that $H$ has property $\M$. Moreover $(\la G\circ f)(x)=\{0\}$ for $x$ off the interior of $K$ and so the property $\M$ of $\la G$ follows from {Fact 2c).}\hfill $\Box$

\bigskip
For the last results of this section we introduce notation which is relevant also to the further sections.
That is, if $A$ is a finite set then we let $\Delta (A)$ denote the $(|A|-1)$--simplex spanned by $A$:
$$\Delta (A):=\{f: A\to [0,1]~|~ \sum _{a\in A} f(a)=1\}$$
and we consider it being embedded in the $|A|$--space $\R^A$ of all functions from $A$ to $\R$, equipped with the norm $\| f\|:=(\sum _a|f(a)|^2)^{1/2}$.

\vskip.2cm Now, $(A_i)_{i\in N}$ be a partition of $A$ into non-empty subsets. Corresponding to it we let $\oDelta : = \prod_{i\in N}  \Delta (A_i)$, considered as a subset of $\prod _{i\in N}\R ^{A_i}=\R ^A$. For $a\in A$ and an element $v$ of $\R^A$ (and hence also for $v\in \oDelta$) we denote by $v^a$ the $a$-th coordinate~of~$v$.

\bigskip
\noindent

\noi {\bf Theorem 4.} {\it
 {Let $F: W\times \oDelta\to \R^A$ be a correspondence with property~$\M$, where $W$ is compact manifold or} is as  in the Addendum to Theorem 1. Let further $D\subset \oDelta\times \R ^A$ denote the set of all pairs $(\sigma, v)$ such that
\begin{equation}\label{*}
\mbox{for all }i\in N\mbox{ and }a\in A_i\mbox{, if }\sigma^a> 0\mbox{ then }v^a=\max \{v^b: b\in A_i\}.
\end{equation}\label{eq1}
Then, $F\cap (W \times D)$ has property~$\M$ for $W$.
 }

\bigskip\noindent {\bf Corollary 3.} {\it  {With $W$ and $F$ as above} define $E: W \to \R^N$  by $(w,r) \in E$ if and only if for some $(w, \sigma, v) \in F$ one has both  (1) above and $r^i = \max_{a \in A_i} v^a$ for each $i \in N$.   Then, $E$ has property $\M$. }

\medskip\noi{\it Proof.} Theorems 2 and 4 do the job, because $E=f\circ (F\cap (W\times D))$ where $f: \R ^A\to \R ^N$ is defined by the formula $f(v)=(\max \{v^a~|~a\in A_i\})_{i\in N}$.
\hfill $\Box$ \vskip.3cm

\noi{\sl
{Proof of Theorem 4.}} Let $G$ denote the closure of $D$ in $\oDelta\times S$, where~$S$~is the sphere $\R ^A\cup \{\infty\}$. By Theorem 1 and its Addendum, combined with the contractibility of $\oDelta$ and of $\R ^A$, it suffices to prove that $G$ has property~$\M$~for~$S$.

To this end let us recall that there exists a retraction $r: \R ^A\to \oDelta$ such that, for $\s\in \oDelta $ and $v\in \R ^A$, condition (1) holds true if and only if $r(\s +v)=\s$. (See (Simon, Spie\.z, Toru\'nczyk (2020), Lemma 1). This $r$ is the product of nearest--point retractions $r_i: \R ^{A_i}\to \Delta (A_i),~i\in N$.) The formulas $\phi (\s, v)=\s+v$ and $\psi (v)=(r(v), v-r(v))$
 hence define mutually inverse mappings $\psi: \R ^A\to D$ and $\phi : D\to\R^A$.
It follows that $\phi: D\to \R ^A$ is a homeomorphism which is at finite distance
from the projection  $(\s, v)\mapsto v$. The straight-line homotopy between
these two mappings is thus proper, and accordingly the projection $p: D\cup \{\infty\}\to S$ is homotopic to the homeomorphism $\phi \cup {\rm{id}} _{\{\infty\}}$.

However, the projection of $G$ to $S$ is equal to the composition $p\circ q$, where $q: G\to D\cup \{\infty\}$ is a map that squeezes the set $G_\infty:=G\cap (\oDelta\times \{\infty\})$ to~$\{\infty\}$ and is the identity elsewhere. It is readily seen that $G_\infty=\oDelta\times \{\infty\}$, whence $q$ induces a homology--isomorphism by Vietoris' theorem. Since so did~$p$, also $p\circ q$ induces such an isomorphism.
\hfill $\Box$

\vskip.5cm

\section{Game bushes}

 \noi We have to modify the concept of a finite game tree so that
 the terminal points of the game  contain variable  payoffs and
 also there could be a variety of locations where
  the game starts.   Without first stipulating the payoffs, we  call this
  modification a {\em game bush}. Some may prefer the term ``game forest''. But our structure involves
  an informational interaction between the different starting points (roots)  and terminal points (leaves); it is not merely an independent collection of games. After we
   add a payoff structure, it becomes a ``game bundle''.
  In Simon, Spie\.z, and Toru\'nczyk (2020) the concept of a ``truncated game tree'' was introduced, a game tree with a unique
  starting point and an informational structure to the terminal points.  However we
  prefer  a  ``game bush'' to
  a ``truncated game bundle'' and prefer a ``subgame bush''
  to a ``truncated subgame bundle''. Except for the terminology,
   the concepts follow in parallel those of that article.
   \vskip.2cm

\noi
 Let there be a  finite set $N$
 of  players.
 There is a finite directed   graph  $(V,V)$ (arrows between  vertices)
  such that the directed
  graph is acyclic  without the orientation of the edges (meaning
  that for every vertex there is only one path leading to that vertex from
   the roots.  The  subset $R$ of
  initial vertices is called the roots.  $T$ is the set of terminal points
     and every path of arrows
    starts at a root and ends at a terminal point,
 with  each terminal point determining
    a unique such path of arrows.
      We allow $T$ and $R$ to have a non-empty
     intersection, meaning that in part of the game bush
     the game could  start and end simultaneously.
       The set $D$ of nodes  is   the  subset
 $ V\backslash T$ and these are the vertices (except for the roots) to which comes exactly one arrow
    and from which, without loss of generality,  come at least two
    arrows. Toward
    a root, there is no arrow, and  from a root comes  at least
     two arrows.
\vskip.2cm

\noi
For  each player $n\in N$  there is a subset $D_n\subseteq D$ such that
$\forall i \not= n\ D_i \cap D_n = \emptyset$.
 Define $D_0$ to be
 the set $D\backslash ( \cup_{n\in N} D_n) $. Also for every $n\in N$ there is
 a partition  ${\cal P}_n$ of $D_n$.  For
 every
 $W\in {\cal P}_n$   there is a  set of actions
 $A_W^n$ such that there is a bijective relationship between $A^n_W$ and the
  arrows leaving every $v\in W$.
 For every $v\in D_0$ there is a probability distribution
 $p_v$ on the arrows leaving the node   $v$, and therefore also
 on the nodes following directly after $v$ in the tree.
 We assume without loss of generality that $p_v$ gives positive probability
 to every arrow leaving $v$. The subset $D_0$ is where
 Nature makes a choice. Nature
 is not a player because it has no payoffs and its actions are involuntary and
  randomized rather than choices in the usual sense.

 \vskip.2cm
\noi   At any node $v\in W \in {\cal
  P}_n$ only the player $n$ is making any decision, and this decision
  determines completely which vertex follows $v$.  At the nodes $v$
 in $V_0$ nature is making a random choice, according to $p_v$,  concerning
  which vertex follows $v$. If the game is at the node $v\in D_n$ and
  $v\in W \in {\cal P}_n$ then Player $n$ is informed   that the node is
  in the set $W$ and that player has no additional information, so that
 inside $W$ player $n$ cannot distinguish between nodes within $W$.
   A set $W\in {\cal P}_n$ is called an {\em information set} of Player $n$.     \vskip.2cm

   \noi The inspiration behind its definition is   that
    a game bush could be
   part of a larger game, either at the start, the end, or in the middle.
   With conventional game trees, there is only one root.
   The introduction of multiple roots
   creates new problems.
   As a game bush could  represent a kind of  subgame,
    information may be inherited from  previous play. Therefore
    for each player $n\in N$ we require that there
     is a partition ${\cal  R}_n$
    of the roots $R$ representing what each  player
    knows at the start of the game bush. And then  to  define the payoffs and
     potentially to connect the game bush to further play in subsequent game bushes,
     we require for each player $n\in N$ that there is a partition
      ${\cal Q}_n$ of the terminal points.
\vskip.2cm

\noi Now we extend the definition of a game bush to
 that of a game bundle. \vskip.2cm

  \noi Define ${\cal Q}$ to be the meet partition
  $\wedge_{n\in N} {\cal Q}_n$, the
  unique finest partition such that for every $n\in N$  every member of
   ${\cal Q}_n$ is contained in some member of ${\cal Q}$.
 For any $C\in {\cal Q}$ we assume
  there is a  correspondence
 $F_C \subseteq \Delta (C) \times {\bf R}^{C\times N}$
  of {\em continuation} payoffs.
  The correspondences
  $F_C$ for all $C\in {\cal Q}$
   together with the
   game bush define a game bundle. The partition
    ${\cal Q}$ represents common knowledge in the end ponts. We believe that the payoffs
   must be defined by such a     partition. A
    non-empty overlap of different  subsets
      could create  an unresolvable
     ambiguity  concerning
     what are   the payoffs.  And an overlap  between an information set  and a set defining the subgame
       would  create problems for  the subgame concept.
\vskip.2cm

\noi    A game
   bundle is not a single game, rather  a  collection of games
   parametrized by the  distributions on the set $R$ of  roots. Furthermore,
   because the payoffs are defined by correspondences,
    there could be different payoff
   consequences for the same set of actions and ending at
   the same terminal point (or indeed none where the correspondence
    may be empty). \vskip.2cm
   \noi
   A game  tree, the conventional game  in extensive form,
   is a game bundle such that
   the set $R$ (of roots)  is a singleton, ${\cal Q}$ is the
   discrete partition on $T$ (meaning the
   collection  $\{ \{  t\} | \ t\in T\} $) and the  correspondence $F_{t} $
    for every $t\in T$ is defined
    by a singleton  $\{ (\delta_t, r)\}$ for some payoff $r\in {\bf R}^N$
    (where $\delta_r$ is the Kroniker delta).
  \vskip.2cm

\noi
 For every player $n\in N$ let  $S_n$ be the finite set of pure decision
 strategies of the players
 in the game bush, by which we mean a function that decides, at every set $W$
 in ${\cal P}_n$  deterministically which
 member of $A^n_W$ should be chosen.
 If each such  $A^n _W$ has cardinality
 $l$ and there are $k$ such sets in ${\cal P}_n$  then the cardinality of
 $S_n$ is $l^k$.    \vskip.2cm
 \noi
 For our purposes, the set $\oDelta
 := \prod_{n\in N} \Delta(S_n)$ will be the set of strategies, what are called
 the mixed strategies. A behaviour strategy is a choice
  of a point in
  $\Delta (A^n_W)$ for each choice of  $n\in N$ and $W\in {\cal P}_n$.
  Later we will establish
  new relations between these different
  types of strategies through  subgame bundles and myopic equilibria.
 \vskip.2cm
\noi
 For every $q\in \Delta (R)$,   $\sigma \in \oDelta := \prod_{n\in N} \Delta(S_n)$,
   and $C\in {\cal Q}$
 let $p_{q,\sigma} (C)$ be
 the probability of reaching $C$ through $\sigma$ and $q\in \Delta (R)$.
 If this probability is positive, define
 $P_{q, \sigma} (\cdot | C)$ to be the  conditional probability on $C$ induced
 by $q$ and $\sigma$.

 \vskip.2cm

  \noi
  Given a game bundle
 with $q\in \Delta (R)$ and $\sigma \in \oDelta $
   define a
   plan $\phi$  for $(q,\sigma)$ to be  a choice of
 $y\in {\bf R} ^{N \times T}$ such that
$y^{N\times C}\in F_C (p_{q,\sigma} ( \cdot | C))  $ if $p_{q,\sigma} (C) >0$
and otherwise $y^{N \times C} \in F_C (w)$ for some $w\in \Delta (C)$.  As
a priori the payoff correspondence may fail to be a non-empty, there
 may be no plan.
\vskip.2cm

Given a  plan $\phi$ for $(q,\sigma)$, a player $n\in N$
  and a choice $s\in S_n$ define
   $f^n_{\phi} (s)$
  to be   the expected value of $y^{n,e}$ as determined by
  the $q\in \Delta (R)$ and transitions
  determined by
  the $s$ and the $(\sigma^j\ | \ j\not=n)$, whereby
  the conditional probabilities $p_{q,\sigma}$ are
 still    determined by the entire  $\sigma = (\sigma^j \ | \ j\in N)$.
  A  plan $\phi$   for $(q, \sigma)$ is an {\em m-equilibrium}   if for all
  $s\in S_n$ with $\sigma^n (s) >0$
   it follows that $f^n _{\phi} (s) = \max_{t\in S_n} f^n_{\phi} (t)$.
   The payoff of a plan $\phi$   is a vector
  in ${\bf R}^N$ such that the $n$ coordinate is the  expected value of
  of the $f^n_{\phi} (s)$, which for an m-equilibrium is the common value
   for the $f_{\phi}^n (s)$ for all those $s$ with positive probability.

  \vskip.2cm
  \noi Next we show that the correspondence of
  m-equilibria has a certain special property as long as
  the correspondences defining the continuation payoffs have that same
  special property.

\medskip

\noi The next theorem applies to the context of a game bundle defined
 by a partition ${\cal Q}$ and correspondences $F_C$ for
 every $C\in {\cal Q}$.

\bigskip
\noi
{\bf Theorem 5.}  {\it If for every $C\in {\cal Q}$ the correspondence
$F_C\subseteq \Delta(C) \times \R^{C\times N}$  has the spanning
property then the  correspondence defined by
 the payoffs of  m-equilibria
 has  the spanning property with respect
 to $\Delta (R)$.}

\medskip
 \noi {\bf Remark:} We follow approximately a previous  argument of
 Simon, Spie{\.z}, and Toru{\'n}czyk (2020).\vskip.2cm

\noi {\sl Proof:} Let  $\epsilon>0$ be given
  and let $B$ be a positive quantity larger than any payoff from
 the correspondences $F_C$.
 If $p_{q,\sigma }(C)$, the probability of reaching $C$ with $q$ and
 $\sigma$, is at least $2\epsilon$ then define  $\lambda_{q,\sigma} (C) =1$.
 If $p_{q,\sigma } (C) \leq \epsilon$ then
  $\lambda_{q,\sigma} (C) =0$.
   And if  $\epsilon < p_{q,\sigma} (C) < 2\epsilon$
   then let   $\lambda_{q,\sigma} (C)=  \frac {p_{q, \sigma}
     (C)-\epsilon} {\epsilon}$.
\vskip.2cm
\noi  For every choice of $\epsilon >0$ and $C\in {\cal Q}$
define  the correspondence
$F_{\epsilon}  \subseteq  \Delta (R) \times \oDelta  \times \R ^{N \times T}$
by $F_{\epsilon} (q, \sigma) ^{N\times C} = \lambda_{q,\sigma} F_C (q,p_{q,\sigma} ( \cdot | C) + (1-\lambda_{q,\sigma}(C) ) \{ r_B\} $,
 where $r_B$ is the point whose entry is $B$ for all choices.
Due to {Corollary~2}
the  correspondence $F_{\epsilon}$ has the spanning property.
Due to {Corollary 3} (with $\Delta(R)$ taking the role of $W$ in Corollary 3), the correspondence $ G_{\epsilon}$ defined by the myopic equilibria from  $F_{\epsilon}$ plans has the spanning property.
And because the spanning property is defined using \v Cech homology,
the cluster  limit $G$  of the $G_{\epsilon}$ for
any sequence of $\epsilon$ going to zero also has the spanning property.
\vskip.2cm
\noi Now consider any  payoff  obtained from $G$;
we need to show that
 it comes from  an m-equilibrium.
 Suppose for a given $(q,\sigma)$
 that $y^{N \times A}$ is from a sequence
  of the $F_{\epsilon}$.
As there are fixed values for the minimal non-zero probability for
reaching any terminal point $e\in T$, and the sequence of $\epsilon$ went
to zero, it  follows for any $C\in {\cal Q}$ given
positive probability by $q$ and $\sigma$
 that $\lambda <1$ was not used in the definition
of the payoffs in $G_{\epsilon}$ for some sufficiently small
$\epsilon$.
And if $C\in {\cal Q}$ is given zero probability by the $q$ and $\sigma$,
due to the
  large positive $B$, any cluster of vectors
  from the  correspondence $F_C$  can be used for the event that
  some terminal point in $C$ is reached
  (and if there is no such cluster, meaning
  that eventually only $B$ was used in the definition of the $G_{\epsilon}$,
  then any vecctor from $F_C$ would do). Given the inequalities defining
  the myopic equilibria of $G$, after removing the use of $B$ those
  inequalities remain to show that $G$ is a subset of the
  myopic equilibria of the game bundle.
  \hfill $\Box$

\section {Subgames and factor games}

\noi   We extend the definition of a subgame from that
defined on a single
 state to one defined on a subset of possible
 states. \vskip.2cm

\noi
{\bf Definition:} A subset $S$ of the vertices $V$ of a game bush
defines a subgame bush  on the vertices $S$ if
and only if \newline
 (1) one cannot leave $S$ through any choice of
 action of a player or a choice of nature (meaning closure by  arrows),
  \newline (2)  for every player $n\in N$ and
 every $W\in {\cal P}_n$ or $W\in {\cal Q}_n$
 either $W$ is  contained in $S$ or is disjoint from $S$.
 \vskip.2cm
 \noi
 At first the condition with ${\cal Q}_n$ may seem unnecessary. But if we  dropped
 this  condition then every subset of $T$ would define a
 subgame bush. That would create a problem if
 the game bush is followed by
 further play.
 \vskip.2cm

\noi  If the vertices
$S$ define a subgame bush, let
$R'$ be the roots of the subgame bush (the vertices
in $S$ that are in $R$ or
whose predecessor is not in $S$) and $T'$ is the subset of
terminal vertices  $T' = T\cap S$.
The partitions ${\cal Q}'_n$ of $T'$ are defined by ${\cal Q}'_n = \{ F\cap S \ | \ F\in {\cal Q}_n\}$
 which by the definition
  of a subgame bush are equal to  $   \{ F\ | \ F\in {\cal Q}_n,
  F \subseteq   T'\}$. Likewise define ${\cal Q}'$ to be the meet partition
 $\wedge_{n\in N} {\cal Q}'_n$.
 The partition ${\cal R}_n'$ of $R'$ is defined
 by $u,v$ belonging to the same member of
 ${\cal R}'_n$ if and only if $u$ and $v$ share  the same  last member of ${\cal P}_n$
 in the paths to $u$ and $v$, and otherwise, if no such member of
 ${\cal P}_n$ exists,
  they share the same member of ${\cal R}_n$ in the paths to $u$ and $v$. \vskip.2cm
  \noi Given that additionally a  game bundle  $\Gamma$ is defined,
 meaning  with  correspondences $F_C$ for every $C\in {\cal Q}$,
   another game bundle is defined by the
  subgame bush defined by $S$ and the  same correspondences $F_C$ for
  the sets $C\in {\cal Q}$, since ${\cal Q}'$ is just a subset of ${\cal Q}$.
  The resulting subgame game bundle  we call
   $\Gamma|_S$, and we see that it is also a game bundle. \vskip.2cm
  \noi Notice that it is possible for a set $S$ defining a subgame bush to have a non-empty intersection
  with the roots $R$. When this happens, the roots $R'$ of the subgame bush have a non-empty intersection with $R$.
  Also it is possible for $R'$ to have a non-empty
  intersection with $T$ (and therefore also with $T'$).

\vskip.2cm
   \noi  For every $S$ that defines a subgame bush, we define  also a  factor game bush.
    Its roots $R''$
    are the same as $R$ and its partitions ${\cal R}_n''$ are the same
     as ${\cal R}_n$.
   Its  terminal points $T''$ are defined to
   be $T'':= (T\backslash S) \cup  R'$, where $R'$ are the roots
    of the subgame bush defined by $S$. The  partition ${\cal Q}_n''$
    of $T''$ are defined by
    $\{ A \ | A \in {\cal Q}_n, A\cap S= \emptyset \} \cup {\cal R}_n'$,
    (where ${\cal R}'_n$ is the partition on the
     roots of the subgame bush). Because $S$ is containing or disjoint
     from every member of ${\cal Q}_n$, the definition of ${\cal Q}_n''$
     is straightforward. We define ${\cal Q}''$ to
      be $\wedge_{n\in N} {\cal Q}_n''$. \vskip.2cm

       \noi {\bf Definition:} A game bush  has {\bf perfect recall}
  for every player $n$ if for  every $v\in  W\in {\cal P}_n$ the member of
    ${\cal R}_n$ followed by the
   information sets in ${\cal P}_n$  and actions taken by
    player $n$ leading to
   $v\in W$ has no repetitions and this sequence is
   the same for every $v\in W$. The game bush
 has perfect recall if it is has perfect recall for every
 player.  \vskip.2cm

 \noi Next we would like to define  a factor game bundle. If
 $C\in {\cal Q}'' \cap {\cal Q}$ is disjoint from $S$,
 we don't change the payoff correspondence
  from the previous definition of $F_C$.
      But we need to know what should be   correspondence
      $F_C$  when $C\in {\cal Q}''$
      is a subset of $ R'$, the roots of the subgame bundle.
     We would like to say that $C$ also defines a subgame bush,
 but unfortunately we cannot do that without the assumption
 of perfect recall. It makes
 sense to define $F_{C}$  to be some  subset of
 the m-equilibria  of the subgame bundle restricted to $\Delta (C)$
 as a subset of $\Delta (R')$.
 \vskip.2cm

 \noi
 If a set $S$ of vertices  defines  a subgame bundle
 then an m-equilibrium for some $q\in \Delta (R)$
 is $S$-perfect if restricted to the
 subgame bundle  defined by $S$
 it defines an m-equilibrium for some distribution on the roots $R'$ of $S$
   such that when
 there is positive probability of reaching $S$ that distribution is
 the conditional probability as implied by the strategies and
  the initial distribution $q$. \vskip.2cm
  An m-equilibrium is subgame bundle perfect if
  it is $S$-perfect for all subsets $S$ of vertices
   that define subgame bundles.
 \vskip.2cm
  \noi

  Unless otherwise stated, for  the factor game bundle
   $\Gamma/ S$ we define the  correspondence
  $F_{C}$ for every $C\in {\cal Q}''$ with $C\subseteq R'$
   to be
   the subgame bundle perfect m-equilibria of the subgame $\Gamma| _S$
    as restricted to $\Delta (C)$.
   If there is some ambiguity
  concerning what should be  $F_{C}$, we can denote it by
 $\Gamma ^F/S$, where $F$ stands for alternative  corespondences.  \vskip.2cm

  \medskip
     \noi
     {\bf Lemma 1.} {\it If the sets $S$ and $T$ define subgame bushes then their union $S\cup T$ and their intersection $S\cap T$ also  define       subgame bushes.   }

     \vskip.2cm
     \noi {\sl Proof:} It follows directly from the definition of subgame bushes as sets  that are closed with
     respect to information sets, partitions,  and consequences of actions.
 \hfill $\Box$    \vskip.2cm
 \noi In general, it may be
     impossible for a player to combine her strategies from $\Gamma/S$ followed by $\Gamma_S$ in
     the way desired, as this player may not possess
     the required memory to do this.
     If we assumed that these were two distinct players, one playing in the factor game and a different one playing in the subgame bundle,
     such a combination would not be problematic. But a  player performing in both the factor game and the subgame
      may fail to
      be capable of performing the necessary mixed strategy in the subgame due to a lack of memory of what happened in the
       factor game.  To go further, we need the property of perfect recall.
 \vskip.2cm

     \noi
     {\bf Lemma {2}:} {\it Let $\Gamma$ be a game bundle,  $S$ a set defining
     the  subgame bundle $\Gamma |_S$ of $G$,
     with $R'$ the roots of the subgame defined by $S$. Assume that the game bush
       of
      $\Gamma$ has perfect recall. Let $\phi$ be
     an m-equilibrium of the factor game $\Gamma/S$ and
     $q$ be its induced conditional probability on $R'$, the roots of $S$.
      The plan $\phi$ combined with any
      m-equilibria of $\Gamma|_S$ corresponding to the conditional
       probability $q$ of $R'$ (with $q$ equal
       to anything if the probability of reaching $S$ is zero)  defines
      an m-equilibrium of $\Gamma$. }

      \vskip.2cm
      \noi {\sl Proof:} Let  $ S^n_1$ and $S^n_2$ be the decision
       functions for $\Gamma/S$ and $\Gamma_S$ respectively.
       Let $s_1$ be a member of $S^n_1$ and  $P$ an information set
       of Player $n$ in the subgame bundle $\Gamma_S$ that is
        reached with positive probability.  If
       $r$ is the expected payoff for Player $n$ from $s_2$
       conditioned on reaching $P$ in the subgame bundle from
       the strategies used in the factor game $\Gamma/S$,
       we need to know that
       $r$ is also the expected payoff from the combination of  $s_1$
        with $s_2$ conditioned on reaching $P$.
        This follows  from
        the property of perfect recall, because the distribution
        on $P$ from using $s_1$ is not any different from the
        distribution on $P$ used to calculate the expected payoff
         $r$ in the subgame bundle.

\bigskip
        \noi {\bf Theorem {6}:} {\it If there is perfect recall, the correspondence of
        subgame bundle perfect m-equilibria has the spanning property, meaning
        that for every intial probability distribution on the roots
        there is a subgame bundle perfect m-equilibrium.}
        \vskip.2cm
        \noi {\sl Proof:} The proof is by induction on the partially ordered
        structure of subgame bundles. If there are no subgame bundles, it
        is true by  {Theorem 1}. Otherwise let $S$ be any subset of vertices
        defining a subgame. By induction we assume that the correspondence
         of subgame bundle perfect m-equilibria   of  both
         $G|_S$ and $G/S$ have the spanning property. Let $U$ be a
         set of vertices defining a subgame. If $U$ is contained in $S$,
         we have already assumed the conclusion. If $U$ is not contained
         in $S$, then  { by Lemma 1}  $G|U$ is a combination of $G|_U / S $ and $G|_{U\cap S}$.
         By  {Lemma 2 and Theorem 2} the subgame perfect m-equilibria
         of  $G|_U$ has  the spanning property.
         \hfill $\Box$         \vskip.2cm

     \section { Further problems}

\noi
The main advantage of mixed strategies over behaviour strategies  is that  one
can define the relationship between strategies and payoffs in a straightforward
 way, through multi-dimensional matrices. However the dimensions of the
 strategy spaces can become very large, and we would like to reduce these dimensions.
\vskip.2cm

\noi {\bf Definition:} An extensive form game is {\em solvable} if there is a
sequence $\emptyset =S_0 \subseteq \cdots \subseteq S_n=V$ such that
for each game $G|_ {S_i} / S_{i-1}$ no player is called upon to make a choice
of action twice (meaning that there are no two $P_1, P_2\in {\cal P}_n$
 such that $P_2$ follows from $P_1$ in some path).\vskip.2cm
 \noi
In a solvable game, in each
factor games  $G|_ {S_i} / S_{i-1}$ there
is no distinction between behaviour strategies and pure strategies.
Therefore one can understand equilibria of the game entirely
 through behaviour strategies using  the above
 theorems. The question however is whether there is any
 computational advantage won through this perspective.
 {So far, the only clear advantage pertains to standard
 games  of incomplete information on one side} where a finite game is followed by an infinitely repeated  one, as
  is done in Simon, Spie\.z, and Toru\'nczyk (2020). {There we proved that there exists  an equilibrium and if the payoffs from both
 games come from the same matrices and the payoff is a combination of the discounted and undiscounted,
 we showed that  $\epsilon$-equilibria exist  for every positive $\epsilon$.  With the above Theorem 5 one could hope to improve this result
 to the existence of an equilibrium. }
\vskip.2cm

\noi   The following examples demonstrate some of the complexity of what is a subgame and what constitutes subgame perfection.
  \vskip.2cm
\noi {\bf Example 2:} \vskip.2cm
   \noi There are  two players. Player One has the choice
    between $A$ for aggression and $P$ for passivity. If $P$ is chosen
    the game moves to State $X$ and      the payoffs are $1$ for Player One
     and $2$ for Player Two.
    If $A$ was chosen the game moves
     to State $Y$ and Player Two has the choice
     between $a$ for aggression and $p$ for passivity. If $a$ is chosen
     the payoff is $0$ for both players and if $p$ is chosen
      the payoff is $2$ for Player One and $1$ for Player Two.
      \vskip.2cm
      \noi There is an equilibrium where Player Two chooses
      $a$ with certainty and  Player One  choose $P$ with certainty. Player
      One is afraid of getting $0$, hence willing to choose $P$.
      And because Player One chooses $P$ with certainty, the ``foolishness''
      of Player Two choosing $a$ never upsets the equilibrium property.
      This is the model of ``mutually assured destruction'', that a player
      promises to harm all players in the game and  that  threat
      works to keep that player from ever confronting the
      situation where that  threat must be carried out.
      On the other hand the
      state $Y$ defines a subgame whose only equilibrium involves
      Player Two choosing $p$, hence also Player One choosing $A$, defining
       a second and very different equilibrium. We
      recongnize the first equilibrium does not induce an equilibrium
      of the subgame defined by the state $Y$, but the second
       equilibrium does. The first equilibrium is subgame perfect, the second is not. \vskip.2cm

\noi But before we discared the second equilibrium as irrelevant, notice that it gives a superior payoff to Player One, the player who is performing
 the interior action in the subgame. \vskip.2cm

       \noi {\bf Example 3:} Now we  alter this game slightly so that         Player Two cannot distinguish
       between State $X$ and State $Y$ and has the actions
       $a$ and $p$ also in State $X$ and the  payoff consequences of
       both $a$ and $p$ from State $X$ are the same as before,
       a payoff of $2$ to Player One and
       $1$ to Player Two. Conventionally, this game has
       no subgames, so every equilibrium is subgame perfect. In some way it is strategically
       equivalent to the previous game, as it can be represented in normal
       form by the same matrix
       $$\begin {matrix}
         & p & a \cr P& (1,2) & (1,2) \cr A & (2,1) & (0,0)    \end{matrix}.$$
       We perceive
        $X$ and $Y$ together as a subset defining  a subgame bundle, with a certainty
        that the play will reach this subgame bundle. Both
        equilibria discovered
         in the previous example,
         $A$ with  $p$ and
           $P$ with  $a$,   induce  equilibria of the
        subgame bundle defined by the set  $\{ X, Y\}$. We introduced a new collection of subgames, and yet the $A$ and $p$ combination cannot be eliminated as a
subgame bundle perfect equilibrium. This shows that the subgame bundle perfect concept is not determined exclusively by the normal, or matrix,  form of the game.

\vskip.2cm

  \noi  One would like a concept of subgame bundle perfection that
  incorporates the most likely probability distribution on a subset
  $C\in {\cal Q}$ given that  the probability for $C$ is zero.
  With probability theory, this question motivates
  the concept of a random variable
   representing conditional probability with the conditioned set has zero probability. However with
   games, especially
   if the game tree is finite,
    this question cannot be  divorced from the related  question
  of who deviated to bring the play to this forbidden subset and why.
 Unfortunately hitherto there is no good general answer to this question.
    \vskip.2cm

\section  {References}

\begin{description}

\medskip

\item Dold, A. (1972), {\em Lectures on Algebraic Topology}, Springer Verlag.

\medskip
\item Kohlberg, E. and Mertens, J.-F. (1986),
On the Strategic Stability
 of Equilibria, {\it Econometrica}, {\bf 54 (5)},  pp. 1003 --1037.

\medskip\item
Kirby, R. C.  and Siebenmann, L. C.  (1977), {\it Foundational essays on topological manifolds, smoothings, and
triangulations}, Annals of Mathematics Studies 88 (Princeton University Press)

  \medskip
 \item Kuhn, H. (1953),  Extensive Games and the Problem
 of Information, in {\em Contributions to the Theory of Games I}, Princeton
 University Press, eds.
 Kuhn and Tucker, pp. 193-216.



\medskip

\item   Simon, R.S., Spie{\.z}, S., Toru{\'n}czyk, H. (2002),
Equilibrium Existence and Topology in Games of Incomplete
Information on
 One Side,   {\it Transactions
 of the American Mathematical Society}, Vol. 354, No. 12,
pp. 5005-5026. \medskip

\item   Simon, R.S., Spie{\.z}, S., Toru{\'n}czyk, H. (2020),
  Games of Incomplete Information and Myopic Equilibria,
  {\it Israel Journal of Mathematics}, to appear.\medskip
\end{description}

\end{document}